\documentclass[11pt]{article}

\usepackage[a4paper,margin=1in]{geometry}
\usepackage{amsfonts,amssymb,amsmath,amsthm}
\usepackage{graphicx}
\usepackage{psfrag}
\usepackage{calc}
\usepackage{epic}
\usepackage[english]{babel}
\usepackage{hyperref}
\usepackage{float}
\usepackage{latexsym}
\usepackage{color}
\usepackage{subfigure,dsfont}

\newtheorem{theorem}{Theorem}

\newtheorem{assumption}[theorem]{Assumption}

\newcommand{\be}{ \begin{equation}}
\newcommand{\ee}{\end{equation}}
\newcommand{\ben}{ \begin{equation*}}
\newcommand{\een}{\end{equation*}}

\def\Rep#1{R^{#1}}
\def\Cod#1{C^{#1}}
\def\Codone{V}

\def\P{{\mathbb P}}

\def\FT{{\overline{F}_T}}
\def\FW{{\overline{F}_W}}

\newcommand{\footremember}[2]{%
    \footnote{#2}
    \newcounter{#1}
    \setcounter{#1}{\value{footnote}}%
}

\title{MDS coding is better than replication for job completion times}
\date{\today}

\author{Ken R. Duffy\footremember{HI}{Hamilton Institute, Maynooth University, Maynooth, Ireland} \and Seva Shneer\footremember{HW}{School of Mathematical and Computer Sciences, Heriot-Watt University, Edinburgh, UK}}

\begin{document}
\maketitle

\begin{abstract}

In a multi-server system, how can one get better performance than
random assignment of jobs to servers if queue-states cannot be
queried by the dispatcher? A replication strategy has recently been
proposed where $d$ copies of each arriving job are sent to servers
chosen at random. The job's completion time is the first time that
the service of any of its copies is complete. On completion, redundant
copies of the job are removed from other queues so as not to
overburden the system.

For digital jobs, where the objects to be served can be algebraically
manipulated, and for servers whose output is a linear function of
their input, here we consider an alternate strategy: Maximum Distance
Separable (MDS) codes. For every batch of $n$ digital jobs that
arrive, $n+m$ linear combinations are created over the reals or a
large finite field, and each coded job is sent to a random server.
The batch completion time is the first time that any $n$ of the
$n+m$ coded jobs are served, as the evaluation of $n$ original jobs
can be recovered by Gaussian elimination. If redundant jobs can be
removed from queues on batch completion, we establish that in order
to get the improved response-time performance of sending $d$ copies
of each of $n$ jobs via the replication strategy, with the MDS
methodology it suffices to send $n+d$ jobs. That is, while replication
is multiplicative, MDS is linear.
\end{abstract}

\section{Introduction}
It is well known that if a job arrives to a system with many servers,
its delay is minimized by joining the queue with the least waiting
time. If there are large numbers of servers, the state of each of
their queues may not be accessible at each job's arrival time. The
celebrated power-of-$d$ choices result (see \cite{azar1999balanced},
\cite{mitzenmacher2001power}, \cite{VDK1996}) establishes that by
sampling a relatively small number of queues at random and joining
the shortest of those, performance is asymptotically much better
than in the case of random assignment. Many other load-balancing
schemes have been proposed and investigated, see \cite{Boor2018}
for a current survey.

An interesting variant of this system has recently been introduced
where the waiting times at any of the queues may not be available
to an arriving job, see \cite{shah2015redundant,gardner2015reducing,MorOR17} and references therein. In
that setting, as opposed to sampling $d$ queues and joining the
shortest waiting time, instead a {\bf replication-d} strategy is
proposed: for each job that arrives, a copy of it is placed in $d$
distinct queues whose lengths and waiting times are unknown. The
job's work is complete when the first of its replicas exits service,
and the remaining $d-1$ duplicate jobs are then removed from the
system. For a heuristic illustration of the performance gain that
is obtained from this strategy, assume that a system is in stationarity
where any job sent to any server experiences a sojourn time, say $W$, comprised
of the waiting time in the queue and the service time once it reaches
the server, with distribution $P(W\leq t)=F_W(t)$. Assuming queues are
independent and identically distributed, a single job arrives and
is subject to replication-d.  With $R$ denoting the job's completition
time, its distribution is given by the minimum
of $d$ independent sojourn times
\begin{align*}
P(\Rep{1,d}>t)=P(\min(W_1,\ldots,W_d)>t)=(1-F_W(t))^{d}. 
\end{align*} 
With the system in stationarity and a batch of $n$ jobs arriving where
each job is subject to replication-d, the tail of the batch completion time
distribution is governed by last job completion time in the batch
and satisfies
\begin{align} 
P(\Rep{n,d}>t) = P(\max(\Rep{1,d}_1,\ldots,\Rep{1,d}_n)>t)
	= 1-\left(1-(1-F_W(t))^{d}\right)^n \sim n (1-F_W(t))^{d},
\label{eq:replication} \end{align}
for large $t$. Thus, through the use of the replication-d strategy,
the tail of the completion time distribution of the batch of $n$
jobs is exponentially curtailed from approximately $n(1-F_W(t))$.
That tail reduction is significant, and greatly reduces the straggler
problem where the system is held up waiting for one job to be served
before it can proceed to the next task. 

In the present paper we consider the performance of an alternative
approach that is available when the jobs to be served can be subject
to algebraic manipulation and the output of the servers is a linear
function of their input. This is the case, for example, if the
jobs consist of digital packets that are traversing a network where
the output of a server is its input and for large matrix multiplication
tasks for Machine Learning. In
the network setting, the replication-d strategy is similar in spirit to
repetition coding \cite{Roth2006}, which is known to be sub-optimal,
and instead we consider a {\bf MDS} (Maximum Distance Separable) approach.
The benefits of MDS codes for making communications robust in
networks subject to packet erasures are well-established. For information
retrevial from a multi-server storage system, the improvements in
response time that is attainable through the use of coding have
been studied 
\cite{huang2012codes,shah2014mds,joshi2014delay,Li2016,lee2017mds}.
Both replication and MDS coding have also been proposed recently
to resolve the stragglers problem in distributed gradient descent
for Machine Learning
\cite{shah2015redundant,lee2017speeding,tandon2017gradient,chen2018draco,halbawi2018improving,maity2018robust,ozfaturay2098speeding,ozfatura2019gradient}.
To the best of our knowledge, however, this is one of the first
times its utility in reducing queueing delay in a feed-forward
system has been shown.

Consider a batch arrival of $n$ jobs, $J_1,\ldots,J_n$, each of
which consists of data of fixed size whose symbols take values
in the reals or a large Galois field. The principle of the MDS
coding approach is that rather than send duplicate jobs, one instead
creates $n+m$ linear combinations of the form
\begin{align*}
K_j = A^{(j)}_1J_1+\cdots+A^{(j)}_nJ_n, \qquad j\in\{1,\ldots,n+m\},
\end{align*}
where the $A^{(j)}_i$ are chosen in the reals or a finite fields.
The principle behind MDS codes is to consider each coded job, $K_j$,
as a random linear equation such that the reciept of any linear
function of any $n$ of the $n+m$ linear combinations allows recovery
of the processing of the original $n$ jobs by Gaussian elimination.
Reed-Solomon codes, for example, \cite{reed1960polynomial} are MDS
codes. More generally, a Random Linear Code, where the coefficients
are chosen uniformly at random, is an MDS code with high probability
for a sufficiently large field size, e.g. \cite{ho2006random}.

When MDS is employed, the completion time of a batch is equal to the
job completion time of any $n$ out of the $n+m$ coded jobs. To
heuristically understand the gain that can be obtained by MDS, again
assume that the system is in stationarity with each queue independently
having a sojourn time distribution $F_W$. A batch of $n$ jobs arrives
and are coded into $n+m$ MDS jobs. Their completition time 
has the same distribution as $n$-th order statistics of $n+m$
random variables with distribution $F_W$, whose complementary 
distribution is known to be given by
\begin{align*}
P(\Cod{n,m}>t) 
%&= P\left((\nmin(W_1,\ldots,W_{n+m})\right>t) \\
& = 
(n+m) {{n+m-1}\choose{n-1}} \sum_{k=0}^{n-1} {{n-1}\choose{k}} (-1)^k \frac{1}{m+k+1} \left(1-F_W(t)\right)^{m+k+1}.
\end{align*}
%where $\nmin(x_1,...,x_{n+m})$ is the $n^\text{th}$ minimum element of the list. 
As $t \to \infty$, the tail is equivalent to its leading term
\begin{align}
(n+m) {{n+m-1}\choose{n-1}} \frac{1}{m+1} \left(1-F_W(t)\right)^{m+1}.
\label{eq:MDS}
\end{align}
Thus the tail of the response time with the MDS strategy is smaller
than the tail of the response time achieved by replication-d in
\eqref{eq:replication} so long as $m \geq d$. This non-rigorous
sketch illustrates the main message of this paper: where $nd$ copies
of jobs are used for replication-d, for digital data subject to
linear processing, MDS can provide better tail response times with only
$n+d$ copies. 

We present the precise model considered in the paper in Section \ref{sec:model}. In Section \ref{sec:with}
we consider the case where $k$, the number of servers, tends to
infinity. Under the mean-field assumption used in the replication-$d$
literature, we demonstrate that as long as $m\ge d$ the tail
distribution of batch completion times of jobs in stationarity is
strictly smaller in the case of MDS when compared with replication-$d$, making the above heuristic arguments rigorous.

%We also present numerical evidence that for values of $m$ slightly
%larger than $d$, the entire tail distribution of the sojourn time
%in MDS is better than in replication-d.

\section{A more precise model} \label{sec:model}

In the rest of the paper we
shall assume that there are $k$ servers, each with an infinite-buffer
queue to store outstanding jobs. Each arrival is a batch of $n$
jobs that appears according to a Poisson process of intensity $\lambda
k/n$, so that, on average, there are $\lambda k$ jobs arriving per
unit of time. For digital data as in communication networks, the
batch arrival assumption is not restrictive as individual jobs can
be sub-divided.  Batch arrivals are also appropriate to represent
the parallelisation of MapReduce computations and more general
parallel-processing computer systems (see, e.g., \cite{Ying2015}).

We assume each version of any job takes an exponential time with
rate $1$ to complete on any server, taken independently of everything
else, including other copies of the same job, and each server's output is a
linear function of its input. 

Another key question is how, once one copy of a job has been processed
to completion, its remaining copies are treated. In some circumstances,
such as the queueing of data jobs in a communications network,
it is not practical to remove copied jobs and they must be served.
In other instances, such as for parallelisation of MapReduce computations,
it would be possible to remove waiting tasks from queues and cease
the service of copies being processed. This latter setting, considered in
\cite{MorOR17} and references therein, provides a model of greater
mathematical interest, and we focus on it in this paper.

\section{MDS vs replication-d with redundant removals} \label{sec:with}

The tails of batch completition times are challenging to analyse, but one can examine the
behaviour in the limit as the number of queues, $k$, becomes
large.

In the system with replication, the job completition time distribution
is derived in \cite[Section 5]{MorOR17}) under an assumption on
asymptotic independence of queues (Assumption \ref{ass:1}, given
below).  It is straightforward to adapt that derivation to the case
of batch arrivals considered here. The tail of the completion time of
any one of the $n$ jobs is given by
\begin{align*}
P(\Rep{1,d}>t) = \left(\frac{1}{\lambda + (1-\lambda) e^{t(d-1)}} \right)^{d/(d-1)}.
\end{align*}
The completion time for $n$ jobs in a batch is then the maximum of $n$
independent random variables with this distribution, and a batch's
response time then has the tail given by
\begin{align} \label{eq:exact_delete_redundancy}
P(\Rep{n,d}>t) =
1-\left(1-\left(\frac{1}{\lambda + (1-\lambda) e^{t(d-1)}} \right)^{d/(d-1)}\right)^n.
\end{align}

Following \cite{MorOR17}, we adopt Assumption \ref{ass:1} on
the asymptotic independence of the queues for our analysis of the
MDS strategy, and refer to that article for a discussion of it.
\begin{assumption}
\label{ass:1}
Let $T_i$ denote the completition time of a job, not subject to removal,
at queue $i$ out of a total of $k$. For $k$ sufficiently large,
the random variables $(T_{i_1}, \ldots T_{i_{n+m}})$ are independent
for any distinct $i_1, \ldots, i_{n+m}$.
\end{assumption}

We prove the following along the lines of the proof introduced in \cite[Section 5]{MorOR17}. We note that a differential equation on the completion times may also be obtained from the result of \cite[Theorem 5.2]{hellemans2019performance} where much more general workload-based policies are considered (using derivations similar to those in Section 6.1 therein). We however present below a simple derivation of \eqref{eq:diff_Equation}, using only straightforward queueing arguments.

\begin{theorem} \label{thm:coding_delete}
For every batch arrival of $n$ jobs, $n+m$ coded jobs are sent to
$n+m$ queues chosen uniformly at random without replacement. As
soon as any $n$ coded jobs are completed, the remaining $m$ jobs
are removed from the system, including those currently in service.
Let $\Codone$ be the random waiting time of a single (virtual) job
that is subject to neither coding or removal, which is needed to
determine the batch waiting time. Under Assumption 1,
its waiting time satisfies the following differential equation 
\begin{align} 
\label{eq:diff_Equation}
& \frac{dP(\Codone>t)}{dt} 
 = - P(\Codone>t) \\ \nonumber
 &+  \alpha(n+m-1) {{n+m-2}\choose{n-1}} \sum_{i=0}^{n-1} {{n-1}\choose{i}} (-1)^i \frac{1}{(m+i)(m+i+1)} (P(\Codone>t))^{m+i+1},
\end{align}
where $\alpha = \lambda(m+n)/n$. For large waiting times, the tail of
its distribution satisfies 
\begin{align*}
\limsup_{t \to \infty}e^t P(\Codone>t) < \infty.
\end{align*}
Let $\Cod{n,m}$ denote the random MDS batch completion time. Its distribution satisfies
\begin{align} \label{eq:response_coding_delete}
\P(\Cod{n,m}>t) = 
(n+m) {{n+m-1}\choose{n-1}} \sum_{i=0}^{n-1} {{n-1}\choose{i}} (-1)^i \frac{1}{m+i+1} \left(P(\Codone>t)\right)^{m+i+1},
\end{align}
and its tail satisfies
\begin{equation} \label{eq:asymp_delete_coding}
\lim\sup_{t \to \infty} e^{(m+1)t} P(\Cod{n,m} > t) < \infty.
\end{equation}
\end{theorem}

We note that this result encompasses the replication-d strategy by
setting $n=1$ and $m=d-1$, and that in that setting
\eqref{eq:diff_Equation} is exactly the differential equation
obtained in \cite{MorOR17} (see the displayed equation just after (24) therein).
These results confirm the heuristic analysis in the introduction
that the tail of the batch completition time distribution for
replication in equation \eqref{eq:exact_delete_redundancy} is slower
than when MDS is used, equation \eqref{eq:asymp_delete_coding},
so long as $m\geq d$. Thus with only $n+d$ coded jobs, one can
achieve better tail performance than sending $nd$ jobs under the
replication strategy.

In the case of replication, a closed form for the batch completition
time distribution is available, but that is not the case for MDS
as the differential equation \eqref{eq:diff_Equation} describing
the virtual waiting time distribution of a non-coded job cannot be
solved in closed form in general. It can, however, be readily solved numerically
and inserted into equation \eqref{eq:response_coding_delete} to
evaluate the batch completition time distribution for MDS. 

\begin{figure}
\centering
\includegraphics[width=0.7\linewidth]{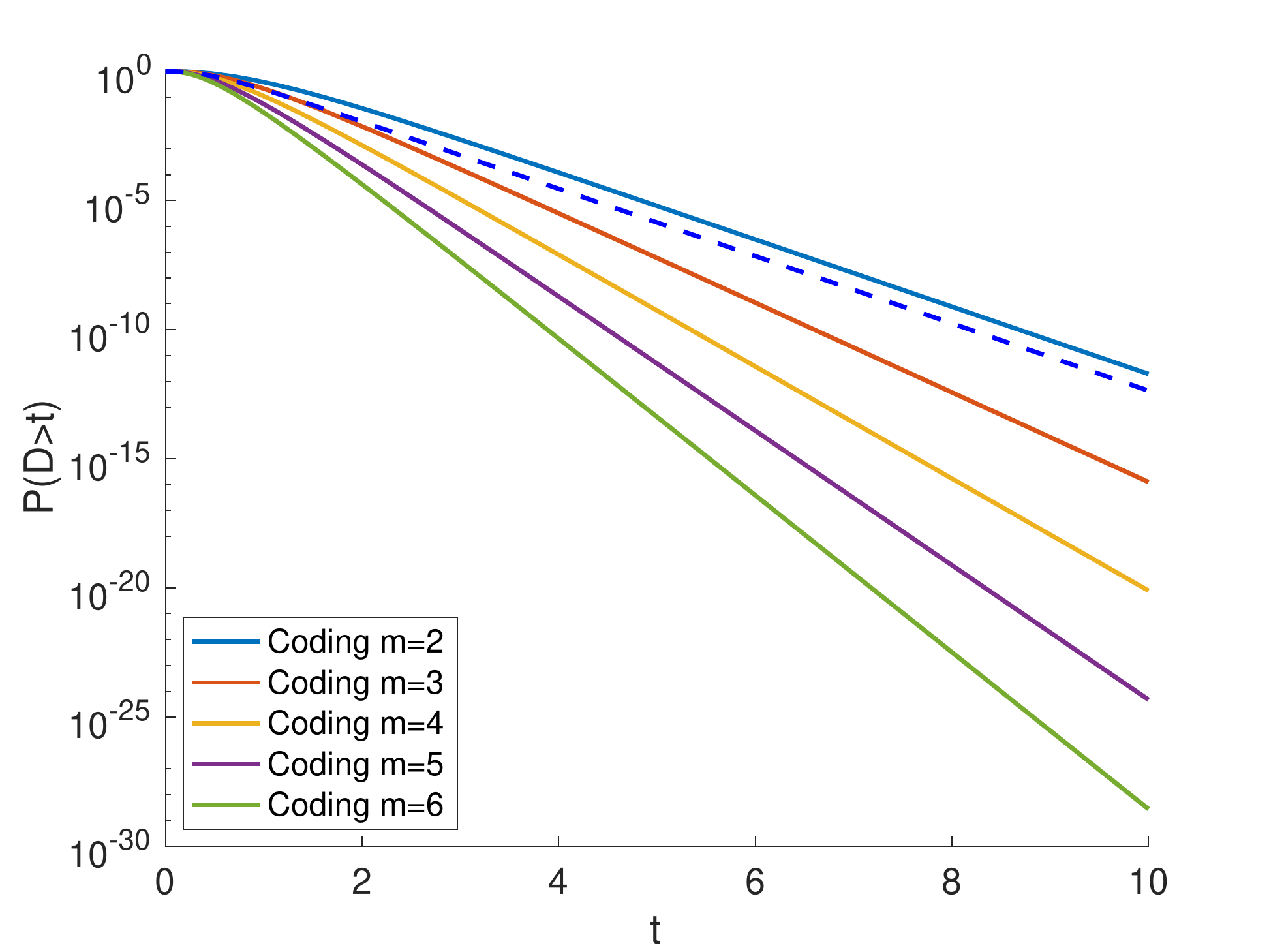}
\caption{Complementary batch completition time distribution for a
batch of $n=3$ jobs. The dashed line corresponds to replication
with $d=3$ giving $nd$ replicated jobs, while the solid lines
correspond to MDS with $n+m$ coded jobs.}
\label{fig:removals_n=3_d=3}
\end{figure}

An example comparison is presented in Fig. \ref{fig:removals_n=3_d=3}
where batches consist of $n=3$ jobs and the replication strategy
places $d=3$ copies of each into the system. For MDS, we consider
a range of values for $m$ from $2$ to $6$. The figure recapitulates
the conclusion that MDS with $m \ge d$ leads to significant gains
in completition time tail for larger values of $t$. Note that for
$m=d=3$, the batch completition time of MDS is not stochastically
dominated by that of replication and that short delays are more
likely with MDS, and it is only in the tail that MDS outperforms
replication. For values of $m\geq4$, however, the MDS batch
completition time distribution is better for all times.

{\bf Proof of Theorem \ref{thm:coding_delete}.} Equation \eqref{eq:response_coding_delete} is a direct application of known results on order-statistics distributions. We now prove \eqref{eq:diff_Equation}. Following derivations in \cite[Section 5]{MorOR17}, denote by $T_i$ the non-redundant response time for a tagged job in queue $i$. Then
$$
T_i = W_i + E_i,
$$
where $W_i$ is the workload (real workload, i.e. time to empty the queue if there were no more arrivals) and $E_i$ is an Exponential$(1)$ random variable (service requirement). Let us denote by $F_T$ and $F_W$ the distribution functions of $T$ and $W$, respectively. Let us also denote their tails by $\FT$ and $\FW$. Then we can write
$$
\FT(t) = e^{-t} + \int_0^t e^{-y} \FW(t-y) dy = e^{-t} + \int_0^t e^{-(t-y)} \FW(y) dy = e^{-t} + e^{-t} \int_0^t e^y \FW(y) dy,
$$
differentiating which we get
\begin{equation} \label{eq:derivative_1}
\FT'(t) = \FW(t) - \FT(t). 
\end{equation}
Let us now write an expression for $\FW(t)$. We can look at the previous arrival (which in the case of MDS happens an Exponential$(\lambda (m+n)/n)$ time earlier than the tagged arrival. For simplicity denote $\alpha = \lambda m/n$. Condition first on the previous arrival having been $y$ time before the tagged arrival. Then $W>t$ in one of two cases: either the previous arrival sees workload larger than $t+y$, or the previous arrival sees workload smaller than $t+y$, its own (non-redundant) time in queue $i$ is larger than $t+y$ and by the time $t+y$ no more than $n-1$ of the other $n+m-1$ copies left other queues (or, in other words, $n$th order statistic of $n+m-1$ random variables exceeds $t+y$);

Integrating over all values of $y$, we obtain:
\begin{align*}
& \FW(t) = \int_0^\infty \alpha e^{-\alpha y} dy \biggl(\FW(t+y) 
\\ & + (\FT(t+y) - \FW(t+y)) (n+m-1) {{n+m-2}\choose{n-1}} \sum_{i=0}^{n-1} {{n-1}\choose{i}} (-1)^i \frac{1}{m+i} (\FT(t+y))^{m+i}
\\ & = e^{\alpha t} \int_t^\infty \alpha e^{-\alpha z} dz \biggl(\FW(z) 
\\ & + (\FT(z) - \FW(z)) (n+m-1) {{n+m-2}\choose{n-1}}  \sum_{i=0}^{n-1} {{n-1}\choose{i}} (-1)^i \frac{1}{m+i} (\FT(z))^{m+i}\biggr) ,
\end{align*}
where we used known results for the tail distribution of $n$th order statistic of $n+m-1$ random variables.
Differentiating the above relation, we get
\begin{align*}
\FW'(t) & = \alpha \FW(t) - \alpha \biggl(\FW(t)
\\ & + (n+m-1) {{n+m-2}\choose{n-1}} (\FT(t) - \FW(t)) \sum_{i=0}^{n-1} {{n-1}\choose{i}} (-1)^i \frac{1}{m+i} (\FT(t))^{m+i}\biggr)
\\ &= \alpha(n+m-1) {{n+m-2}\choose{n-1}} \FT'(t) \sum_{i=0}^{n-1} {{n-1}\choose{i}} (-1)^i \frac{1}{m+i} (\FT(t))^{m+i},
\end{align*}
where in the last equality we used \eqref{eq:derivative_1}. The above can be integrated and plugged into \eqref{eq:derivative_1} to obtain
$$
\FT'(t) = - \FT(t) +  \alpha(n+m-1) {{n+m-2}\choose{n-1}} \sum_{i=0}^{n-1} {{n-1}\choose{i}} (-1)^i \frac{1}{(m+i)(m+i+1)} (\FT(t))^{m+i+1}.
$$
This proves \eqref{eq:diff_Equation}.
Recall that $m-n \ge 1$. Since $\FT(x) \to 0$ as $x \to \infty$, there exists $X$ such that
$$
\FT'(x) \le -\FT(x) + A (\FT(x))^2,
$$
for all $x \ge X$, with some constant $A > 0$. Denote by $g(x) = 1/\FT(x)$. Then
$$
g'(x) = - \frac{\FT(x)}{(\FT(x))^2} \ge - \frac{-\FT(x) + A (\FT(x))^2}{(\FT(x))^2} = g(x) - A
$$
for all $x \ge X$. Since $g(x) \to \infty$ as $x \to \infty$, we can also assume that $g(x) \ge A$ for all $x \ge X$. Then the above implies that
$$
\frac{g'(x)}{g(x) - A} \ge 1
$$
for all $x \ge X$. Integrating the above inequality from $X$ to $x$ implies
$$
\log\left(\frac{g(x)-A}{g(X)-A}\right) \ge x-X,
$$
and hence $g(x) \ge A + (g(X) - A) e^{x-X} \ge B e^{x}$, with a constant $B$. This of course implies \eqref{eq:asymp_delete_coding}, and the proof is complete. \qed

\bibliographystyle{abbrv}
\bibliography{coding_replication}

\end{document}